\documentclass[a4paper,11pt]{article}

\usepackage{amsmath,amssymb}     %%一些数学符号
\usepackage{color}
\usepackage{graphicx}
\usepackage{subfigure}
\usepackage{cite}                %%是引用格式变漂亮  %%PRD等APS模板必须禁用该项
\usepackage{hyperref}            %%超链接必须
\usepackage{multirow,makecell}
\usepackage{braket}   %%表格

\numberwithin{equation}{section}   %%公式按节编号

\newcommand{\matr}[2]{\left(\begin{array}{#1}#2\end{array}\right)}
\def \be {\begin{equation}}
\def \ee {\end{equation}}
\def \ba {\begin{array}}
\def \ea {\end{array}}
\def \bea{\begin{eqnarray}}
\def \eea{\end{eqnarray}}
\def \nn {\nonumber}

\def \a {\alpha}

\def \dg {\dagger}

\def \l {\lambda}

\def \R1 {\uppercase\expandafter{\romannumeral1}}
\def \R2 {\uppercase\expandafter{\romannumeral2}}
\def \R3 {\uppercase\expandafter{\romannumeral3}}

\def \f {\frac}

\def \nn {\nonumber}

\def \lt {\left}
\def \rt {\right}

\def \Tr {{\textrm{Tr}}}

\def \diag {{\textrm{diag}}}

\def \and {{\textrm{and}}}

\begin{document}
\begin{titlepage}
	
	\title{\textbf {Asymptotic Bethe ansatz of ABJM open spin chain from giant graviton}}
	\author{Hui-Huang Chen\footnote{chenhh@jxnu.edu.cn}~,}
	\date{}
	
	\maketitle
	\underline{}
	\vspace{-12mm}
	
	\begin{center}
		{\it
             College of Physics and Communication Electronics, Jiangxi Normal University,\\ Nanchang 330022, China\\
		}
		\vspace{10mm}
	\end{center}
	\begin{abstract}
	  In our previous work, the two-loop integrability of ABJM determinant like operator has been well established. In this paper, we push the integrability to all loop orders. The asymptotic Bethe ansatz equations for ABJM determinant like operator (open string attached on giant graviton) are obtained. In the derivation, the symmetries preserved by the bulk and the boundary played a crucial role. Taking the weak coupling limit and applying appropriate fermionic dualities, we obtained a different set of scalar sector Bethe equations with our previous results. When the ``gauge" transformation on Bethe equations was introduced, the discrepancy disappeared.
	\end{abstract}
	
\end{titlepage}

%\vspace{-10mm}

%\begin{center}
%{\it
%Department of Physics and State Key Laboratory of Nuclear Physics and Technology,\\Peking University, 5 Yiheyuan Road, Beijing 100871, %China%\\
%\vspace{2mm}
%$^{2}$Collaborative Innovation Center of Quantum Matter, 5 Yiheyuan Rd, Beijing 100871, P.~R.~China\\
%$^{3}$Center for High Energy Physics, Peking University, 5 Yiheyuan Rd, Beijing 100871, P.~R.~China
%}
%\vspace{10mm}
%\end{center}

%\begin{abstract}

%\end{abstract}

%\baselineskip 18pt
%\thispagestyle{empty}

%\newpage

%\tableofcontent
\section{Introduction}
Integrability plays a key role in quantitative understanding of the planar $\mathrm{AdS}_5/\mathrm{CFT}_4$ correspondence and strongly coupled quantum field theories (QFT). For a review, see \cite{Beisert:2010jr} and references therein. Exploration of integrability based methods to studying non-perturbative properties of QFT is one of the research focuses in theoretical physics. Integrable quantum field theories in spacetime dimension higher than two are quite rare, among which the planar $\mathcal{N}=4$ super Yang-Mills (SYM) theory in four-dimensional spacetime is the most famous example. \\
\par $\mathcal{N}=6$ superconformal Chern-Simons in three-dimensional spacetime is another example \cite{Aharony:2008ug}. The integrable structure established in the ABJM theory mainly focuses on single trace operators which correspond to closed spin chains in field theory or closed strings in the gravity dual theory \cite{Minahan:2008hf, Gromov:2008qe, Ahn:2008aa, Cavaglia:2014exa}. See review \cite{Klose:2010ki} and references therein. In our previous paper \cite{Chen:2018sbp}, we studied the anomalous dimension problem of determinant like operators in ABJM theory under the inspiration of the fact that integrable structure can be found in SYM open spin chain constructed from giant graviton\cite{Berenstein:2005vf, Hofman:2007xp}.
We have computed the anomalous dimension matrix of the following determinant like operators up to two-loop order
\begin{equation}
O_W = \epsilon_{a_1...a_N}\epsilon^{b_1...b_N}(A_1 B_1)^{a_1}_{b_1}...(A_1 B_1)^{a_{N-1}}_{b_{N-1}}W^{a_N}_{b_N},
\end{equation}
where
\begin{equation}
W=(A_2B_2)\cdots(\chi)\cdots(A_2B_2).
\end{equation}\\
\par The determinant like operators in the ABJM theory can be viewed as an open spin chain with the Hamiltonian given by their anomalous dimension matrix. The gravity dual description of these operators is open strings attached to the giant graviton--D4-brane wrapping a $\mathbb{CP}^2$ inside $\mathbb{CP}^3$, while the operator with $W=A_1B_1$ is dual to the D4-brane itself \cite{Giovannoni:2011pn, Cardona:2014ora}.
Strong evidence on the integrability of this open spin chain has been found in paper \cite{Chen:2018sbp} based on the coordinate Bethe ansatz approach and a much more solid proof was given in \cite{Bai:2019soy}. In this paper, we want to push these results to higher loops to obtain the asymptotic Bethe ansatz equations. We do this by the assumption of integrability still survives at higher loops.
We choose $W=(A_2B_2)^L$ as our open spin chain vacuum, and the field $\chi$ is an elementary excitation belongs to either A-particles or B-particles and both are transformed in the four-dimensional representation of centrally extended $\mathfrak{su}(2|2)$ algebra. At the boundaries, the symmetry preserved by the bulk breaks down to $\mathfrak{su}(1|2)$.\\
\par The ABJM giant graviton open spin chain has two types of particle which we named A-particle and B-particle and are charge conjugate to each other. The bulk dispersion relation is
\be
\epsilon(p)=\frac{1}{2}\sqrt{1+16h^2(\l)\sin^2(\frac{p}{2})}
\ee
where the $h(\l)$ is the so called interpolation function with the weak coupling expansion \cite{Minahan:2009aq, Minahan:2009wg, Leoni:2010tb}
\be
h(\lambda)=\lambda-\frac{\pi^2}{3}\lambda^3+\mathcal{O}(\lambda^5),
\ee
and in paper \cite{Gromov:2014eha}, the authors conjectured a exact formula of $h(\l)$ by comparing the quantum spectral
curve method \cite{Cavaglia:2014exa} and supersymmetric localization.
The anomalous dimension of determinant like operator is related to the bulk energy of the spin chain as
\be
\Delta=\sum_{j=1}^{K^I_A}\left(\epsilon(p_j^A)-\frac{1}{2}\right)+\sum_{j=1}^{K^I_B}\left(\epsilon(p_j^B)-\frac12\right)
\ee
where $K^I_A, K^I_B$ is the number of momentum carrying A-particle and B-particle respectively.\\
\par The remaining part of this paper is organized as follows: In section \ref{sec2}, we briefly review the $\mathfrak{su}(2|2)$ invariant S-matrix and obtain the bulk S-matrices of ABJM giant graviton open spin chain. In section \ref{sec3}, we fix the boundary scattering amplitudes by symmetry analysis and computation from the weak coupling region. In section \ref{sec4}, we define the double row transfer matrices and the eigenvalues are easily obtained based on people's previous work. The asymptotic Bethe ansatz equations are also given in this section. In section \ref{sec5}, we discuss the weak coupling limit of our asymptotic Bethe ansatz equations. By a sequence of actions of fermionic dualities, we obtain the two-loop scalar sector (SU(4) sector) Bethe equations. Making use of the ``gauge" transformation on Bethe equations, we can obtain the same set of equations derived in our previous paper. Finally, we conclude in section \ref{sec6} and some details on ``gauge" transformation are given in the appendix \ref{appenA}.
\section{The bulk S-matrices}\label{sec2}
As mentioned above, the symmetry preserved by the bulk of the ABJM giant graviton open spin chain is centrally extended $\mathfrak{su}(2|2)$.
It's convenient to use the generalized rapidity $z_1,z_2$ to parameterize the $\mathfrak{su}(2|2)$ invariant S-matrix. In this paper, we use the form of $\mathfrak{su}(2|2)$ invariant S-matrix given in \cite{Arutyunov:2008zt}
\be
S(z_1,z_2)=\sum_{k=1}^{10}a_{k}(z_1,z_2)\Lambda_k,
\ee
where $\Lambda_k$ are $\mathfrak{su}(2)\oplus\mathfrak{su}(2)$ invariant matrices and $a_k(z_1,z_2)$ are the corresponding coefficients. The explicit expressions can be found in that paper.
\par It's useful to introduce spectral parameter $x$ and $u$
\be
x^{\pm}+\frac{1}{x^{\pm}}=\frac{u\pm\frac{i}{2}}{h(\l)},\quad x^{\pm}\equiv x(u\pm\frac{i}{2}).
\ee
The momentum $p$ and energy $\epsilon$ of the fundamental magnon are
\be
e^{ip}=\frac{x^+}{x^-},\quad \epsilon=\frac12+ih(\l)(\frac{1}{x^+}-\frac{1}{x^-}).
\ee
The generalized rapidity is defined on a torus with half periods
\be
\omega_1=2K(k),\qquad \omega_2=2iK(1-k)-2K(k),
\ee
where $K(k)$ stands for the complete elliptic integral of the first kind with elliptic modulus $k=-16h^2(\l)$.
The spectral parameters $x$, the momentum and the energy of magnon can be expressed in terms of Jacobi elliptic functions of the generalized rapidity $z$
\be
x^{\pm}=\frac{1}{4h(\l)}(\frac{\mathrm{cn} z}{\mathrm{sn} z}\pm i)(1+\mathrm{dn} z),\quad p(z)=2\mathrm{am}z, \quad \epsilon(z)=\frac12\mathrm{dn}(z).
\ee
\par The S-matrix is unitary
\be
S_{12}(z_1,z_2)S_{21}(z_2,z_1)=I_{12},
\ee
and satisfies the Yang-Baxter equation
\be
S_{12}(z_1,z_2)S_{13}(z_1,z_3)S_{23}(z_2,z_3)=S_{23}(z_2,z_3)S_{13}(z_1,z_3)S_{12}(z_1,z_2).
\ee
Here $I$ is the identity matrix.\\
\par The particle anti-particle transformation or crossing transformation is defined by
\be
p(z)\rightarrow -p(z),\quad \epsilon(z)\rightarrow -\epsilon(z),
\ee
which can be described by the shift of the rapidity along the imaginary half period $\omega_2$
\be
x^{\pm}(z)\rightarrow\frac{1}{x^{\pm}(z)}=x^{\pm}(z\pm\omega_2).
\ee
We also note that the reflection
\be
p(z)\rightarrow -p(z),\quad \epsilon(z)\rightarrow \epsilon(z)
\ee
corresponds to a transformation on rapidity $z\rightarrow -z$
\be
x^{\pm}(-z)=-x^{\mp}(z).
\ee
\par The S-matrix also satisfies the (quasi) crossing relations
\be\label{quisacross}
\begin{split}
&C_1^{-1}S_{12}^{t_1}(z_1,z_2)C_1S_{12}(z_1+\omega_2,z_2)=\frac{1}{f(x_1,x_2)}I_{12},\\
&S_{12}^{t_2}(z_1,z_2)C_2S_{12}(z_1,z_2-\omega_2)C_2^{-1}=\frac{1}{f(x_1,x_2)}I_{12},
\end{split}
\ee
where $C$ is the charge conjugation matrix
\be
C=\matr{cc}
{
\sigma_2&0\\
0&i\sigma_2\\
}.
\ee
Here $\sigma_2$ is the Pauli matrix and $f(x_1,x_2)$ is a scalar function defined by
\be
f(x_1,x_2)=\frac{(x_1^+-x_2^-)(1-\frac{1}{x_1^-x_2^-})}{(x_1^+-x_2^+)(1-\frac{1}{x_1^-x_2^+})}.
\ee
\par The scattering matrices of A-particle and B-particle are given by
\begin{equation}
\begin{split}
\mathbb{S}^{AA}(z_1,z_2)=\mathbb{S}^{BB}(z_1,z_2)=S_0(z_1,z_2)S(z_1,z_2),\\
\mathbb{S}^{AB}(z_1,z_2)=\mathbb{S}^{BA}(z_1,z_2)=\tilde{S}_0(z_1,z_2)S(z_1,z_2).
\end{split}
\end{equation}
We assume $\mathbb{S}^{AA}$ and $\mathbb{S}^{AB}$ satisfy the unitary conditions, which imply
\be\label{unitary}
S_0(z_1,z_2)S_0(z_2,z_2)=1,\qquad \tilde{S}_0(z_1,z_2)\tilde{S}_0(z_2,z_1)=1.
\ee
The identification of the B-particles as charge conjugates of the A-particles suggests the following crossing relations \cite{Ahn:2008aa}
\be
\begin{split}
C_1^{-1}\mathbb{S}_{12}^{AAt_1}(z_1,z_2)C_1\mathbb{S}_{12}^{AB}(z_1+\omega_2,z_2)=I_{12},\\
\mathbb{S}_{12}^{AAt_2}(z_1,z_2)C_2\mathbb{S}_{12}^{AB}(z_1,z_2-\omega_2)C_2^{-1}=I_{12}.
\end{split}
\ee
Then using the relation ~\ref{quisacross}, the scalar factor should satisfy
\be\label{crossing}
S_0(z_1,z_2)\tilde{S}_0(z_1+\omega_2,z_2)=S_0(z_1,z_2)\tilde{S}_0(z_1,z_2-\omega_2)=f(x_1,x_2).
\ee
The constraints eq.~\ref{unitary} and eq.~\ref{crossing} can be solved as\footnote{The equations satisfied by the scalar of S-matrices has another solution, which is given by $S_0(z_1,z_2)$ and $\tilde{S}_0(z_1,z_2)$ interchanged in eq.~\ref{ScalarS}. These two solutions correspond to two possible choices of grading in ABJM higher loops Bethe ansatz equations. We adopt the $\eta=+1$ grading or $\mathfrak{su}(2)$ grading in the main text.}
\be\label{ScalarS}
S_0(z_1,z_2)=\frac{x_1^+-x_2^-}{x_1^--x_2^+}\frac{1-\frac{1}{x_1^+x_2^-}}{1-\frac{1}{x_1^-x_2^+}}\sqrt{\frac{x_1^-}{x_1^+}}\sqrt{\frac{x_2^+}{x_2^-}}\sigma(z_1,z_2), \qquad \tilde{S}_0(z_1,z_2)=\sqrt{\frac{x_1^-}{x_1^+}}\sqrt{\frac{x_2^+}{x_2^-}}\sigma(z_1,z_2),
\ee
where $\sigma$ is the BES dressing phase \cite{Beisert:2006ez} and has the following properties
\be
\begin{split}
&\sigma(z_1,z_2)\sigma(z_2,z_1)=1,\quad \sigma(z_1+\omega_2,z_2)\sigma(z_1,z_2)=\frac{x_2^-}{x_2^+}\frac{(x_1^--x_2^+)(1-\frac{1}{x_1^-x_2^-})}{(x_1^+-x_2^+)(1-\frac{1}{x_1^+x_2^-})},\\
&\sigma(z_1,z_2-\omega_2)\sigma(z_1,z_2)=\frac{x_1^+}{x_1^-}\frac{(x_1^--x_2^-)(1-\frac{1}{x_1^-x_2^+})}{(x_1^+-x_2^+)(1-\frac{1}{x_1^+x_2^-})}.
\end{split}
\ee
\section{Boundary reflection matrices}\label{sec3}
\subsection{Boundary $\mathfrak{su}(1|2)$ symmetry and boundary crossing}
In paper \cite{Chen:2018sbp}, we have computed the reflection matrices of the scalar sector at leading order (two loops). We find the reflection matrices are diagonal which means the A-particle and B-particle will not mix when they are scattered at the boundaries (at least at two-loop level). We assume that this property is still preserved at higher loops. Thus there are two kinds of reflection matrices at each boundary. Similar to the analysis of $\mathcal{N}=4$ SYM open string attached to $Y=0$ brane \cite{Hofman:2007xp}, symmetry preserved by the boundary of our integrable open system is $\mathfrak{su}(1|2)$, which can fix the right boundary reflection matrices up to some scalar factors
\be
\begin{split}
\mathbb{R}^{A-}(p)=R_0^{A-}(p)R^-(p)=R_0^{A-}(p)\diag(e^{-i\frac{p}{2}},-e^{i\frac{p}{2}},1,1),\nn\\ \mathbb{R}^{B-}(p)=R_0^{A-}(p)R^-(p)=R_0^{B-}(p)\diag(e^{-i\frac{p}{2}},-e^{i\frac{p}{2}},1,1).
\end{split}
\ee
The left boundary reflection matrices are related to the right ones as
\be
\mathbb{R}^{A+}(p)=\mathbb{R}^{A-}(-p),\quad \mathbb{R}^{B+}(p)=\mathbb{R}^{B-}(-p).
\ee
There are two types of particles in the bulk, both are transformed under fundamental representation of $\mathfrak{su}(2|2)$.  Therefore the bulk Zamolodchikov-Faddeev (ZF) algebra is described by two kinds of creating operators
\be
\begin{split}
\mathbb{A}^{\dg}_i(p), \mathbb{B}^{\dg}_i(p), \qquad i=1,\cdots,4\nn
\end{split}
\ee
which satisfy
\be
\begin{split}
\mathbb{A}^{\dg}_i(p_1)\mathbb{A}^{\dg}_j(p_2)=\mathbb{S}^{AA}(p_1,p_2)_{ij}^{i'j'}\mathbb{A}^{\dg}_{j'}(p_2)\mathbb{A}^{\dg}_{i'}(p_1),\\
\mathbb{B}^{\dg}_i(p_1)\mathbb{B}^{\dg}_j(p_2)=\mathbb{S}^{BB}(p_1,p_2)_{ij}^{i'j'}\mathbb{B}^{\dg}_{j'}(p_2)\mathbb{B}^{\dg}_{i'}(p_1),\\
\mathbb{A}^{\dg}_i(p_1)\mathbb{B}^{\dg}_j(p_2)=\mathbb{S}^{AB}(p_1,p_2)_{ij}^{i'j'}\mathbb{B}^{\dg}_{j'}(p_2)\mathbb{A}^{\dg}_{i'}(p_1),\\
\mathbb{B}^{\dg}_i(p_1)\mathbb{A}^{\dg}_j(p_2)=\mathbb{S}^{BA}(p_1,p_2)_{ij}^{i'j'}\mathbb{A}^{\dg}_{j'}(p_2)\mathbb{B}^{\dg}_{i'}(p_1).
\end{split}
\ee
The associativity of the bulk ZF algebra implies the Yang-Baxter equation (YBE). In order to incorporate the boundary integrable systems, it's useful to introduce the boundary creating operator $\mathcal{B}$\cite{Ghoshal:1993tm}. Following the strategy in \cite{ Ahn:2008df}, for the right boundary, we introduce the right boundary creating operator $\mathcal{B}_R$ satisfying the right boundary ZF algebra
\be
\begin{split}
\mathbb{A}^{\dg}_i(p)\mathcal{B}_R=\mathbb{R}^{A-}(p)_i^{i'}\mathbb{A}^{\dg}_{i'}(-p)\mathcal{B}_R,\\
\mathbb{B}^{\dg}_i(p)\mathcal{B}_R=\mathbb{R}^{B-}(p)_i^{i'}\mathbb{B}^{\dg}_{i'}(-p)\mathcal{B}_R.
\end{split}
\ee
The consistent condition of the bulk ZF algebra and the right boundary ZF algebra gives the boundary Yang-Baxter equation (BYBE).
It's easy to see the right boundary ZF algebra implies the right boundary unitary
\be
\mathbb{R}^{A-}(p)\mathbb{R}^{A-}(-p)=I,\quad \mathbb{R}^{B-}(p)\mathbb{R}^{B-}(-p)=I.
\ee
In terms of scalar factors, we have
\be\label{Bunitary}
R_0^{A-}(p)R_0^{A}(-p)=1,\quad R_0^{B-}(p)R_0^{B}(-p)=1.
\ee
In order to obtain the boundary crossing unitary relation, we consider the following singlet operator
\be
\mathbb{I}(p)=\mathbb{I}_1(p)+\mathbb{I}_2(p),
\ee
where
\be
\mathbb{I}_1(p)=C^{ij}\mathbb{A}_i^{\dagger}(p)\mathbb{B}_j^{\dagger}(\bar{p}),\quad \mathbb{I}_2(p)=C^{ij}\mathbb{B}_i^{\dagger}(p)\mathbb{A}_j^{\dagger}(\bar{p}),
\ee
and $\bar{p}$ is the crossed momentum, defined by
\be
x^{\pm}(\bar{p})=\frac{1}{x^{\pm}(p)}.
\ee
Scattering the singlet operator off the right boundary, we must have
\be\label{bsinglet}
\mathbb{I}(p)\mathcal{B}_R=\mathbb{I}(-\bar{p})\mathcal{B}_R.
\ee
Firstly, considering the $\mathbb{I}_1(p)$ term, we obtain
\be
\begin{split}
&\mathbb{I}_1(p)\mathcal{B}_R=C^{ij}\mathbb{A}_i^{\dagger}(p)\mathbb{B}_j^{\dagger}(\bar{p})\mathcal{B}_R\\
&=C^{ij}\mathbb{R}^{B-}(\bar{p})^{j'}_j\mathbb{A}_i^{\dagger}(p)\mathbb{B}_{j'}^{\dagger}(-\bar{p})\mathcal{B}_R\\
&=C^{ij}\mathbb{R}^{B-}(\bar{p})^{j'}_j\mathbb{S}^{AB}(p,-\bar{p})_{ij'}^{i'j''}\mathbb{B}_{j''}^{\dagger}(-\bar{p})\mathbb{A}_{i'}^{\dagger}(p)\mathcal{B}_R\\
&=C^{ij}\mathbb{R}^{B-}(\bar{p})^{j'}_j\mathbb{S}^{AB}(p,-\bar{p})_{ij'}^{i'j''}\mathbb{R}^{A-}(p)^{i''}_{i'}\mathbb{B}_{j''}^{\dagger}(-\bar{p})\mathbb{A}_{i''}^{\dagger}(-p)\mathcal{B}_R\\
&\equiv \mathbb{I}_2(-\bar{p})\mathcal{B}_R.
\end{split}
\ee
Similarly, for the $\mathbb{I}_2(p)$ term, we have
\be
\begin{split}
&\mathbb{I}_2(p)\mathcal{B}_R=C^{ij}\mathbb{B}_i^{\dagger}(p)\mathbb{A}_j^{\dagger}(\bar{p})\mathcal{B}_R\\
&=C^{ij}\mathbb{R}^{A-}(\bar{p})^{j'}_j\mathbb{S}^{BA}(p,-\bar{p})_{ij'}^{i'j''}\mathbb{R}^{B-}(p)^{i''}_{i'}\mathbb{A}_{j''}^{\dagger}(-\bar{p})\mathbb{B}_{i''}^{\dagger}(-p)\mathcal{B}_R\\
&\equiv \mathbb{I}_1(-\bar{p})\mathcal{B}_R.
\end{split}
\ee
Using the relation \ref{bsinglet}, we obtain
\be
\begin{split}
C^{ij}\mathbb{R}^{B-}(\bar{p})_j^{j'}(\bar{p})\mathbb{S}^{AB}(p,-\bar{p})_{ij'}^{i'j''}\mathbb{R}^{A-}(p)_{i'}^{i''}=C^{j''i''},\\
C^{ij}\mathbb{R}^{A-}(\bar{p})_j^{j'}(\bar{p})\mathbb{S}^{BA}(p,-\bar{p})_{ij'}^{i'j''}\mathbb{R}^{B-}(p)_{i'}^{i''}=C^{j''i''}.
\end{split}
\ee
In terms of the scalar factors $R^{A-}_0(p),R^{B-}_0(p)$, the above boundary crossing relations imply
\be\label{Bcrossing}
R^{A-}_0(p)R^{B-}_0(\bar{p})=\frac{1}{\sigma(p,-\bar{p})}.
\ee
We define
\be
f_b(p)=\frac{x^-+\frac{1}{x^-}}{x^++\frac{1}{x^+}}.
\ee
A solution of eq.~\ref{Bunitary} and eq.~\ref{Bcrossing} is given by the ansatz\footnote{We have utilized the solution given in \cite{Chen:2007ec}: $\sigma(p,-p)\sigma(\bar{p},-\bar{p})=\frac{f_b(p)}{\sigma^2(p,-\bar{p})}$ and the fact $f_b(p)=f_b(\bar{p}),f_b(p)f_b(-p)=1$.}
\be
R^{A-}_0(p)=R^{B-}_0(p)=R_0^{-}(p),
\ee
where
\be
R_0^{-2}(p)=F(p)\sigma(p,-p)\frac{1}{\sqrt{f_b(p)}}
\ee
and $F(p)$ is a CDD-type factor satisfies
\be
F(p)F(\bar{p})=1,\quad F(p)F(-p)=1.
\ee
The CDD-type factor $F(p)$ can be fixed by comparing with the reflection matrix obtained from the weak coupling result
\be
F(p)=-e^{-\frac{ip}{2}},
\ee
which we will discuss in the next subsection.
\subsection{Fixing the CDD factor}
We now turn to the weak coupling region. We can fix the CDD factor by comparing the quantized momentum of single particle excitation computed in two different ways \cite{Bajnok:2010ui}.
\par The two-loop Hamiltonian of the ABJM open spin chain from giant graviton is given in \cite{Chen:2018sbp}
\begin{equation}
\begin{split}
H=&\lambda^2\sum_{l=2}^{2L-3}\left(\mathbb{I}-\mathbb{P}_{l,l+2}+\frac{1}{2}\mathbb{P}_{l,l+2}\mathbb{K}_{l,l+1}+\frac{1}{2}\mathbb{P}_{l,l+1}\mathbb{K}_{l+1,l+2}\right)Q^{A_1}_1Q^{B_1}_{2L}\\
&+\lambda^2Q_1^{A_1}\left(\mathbb{I}+\frac{1}{2}\mathbb{K}_{1,2}-\mathbb{P}_{1,3}+\frac{1}{2}\mathbb{P}_{1,3}\mathbb{K}_{1,2}+\frac{1}{2}\mathbb{P}_{1,3}\mathbb{K}_{2,3}\right)Q_1^{A_1}Q^{B_1}_{2L}\\
&+\lambda^2Q^{A_1}_1Q^{B_1}_{2L}\left(\mathbb{I}+\frac{1}{2}\mathbb{K}_{2L-1,2L}-\mathbb{P}_{2L-2,2L}+\frac{1}{2}\mathbb{P}_{2l-2,2l}\mathbb{K}_{2L-2,2L-1}
+\frac{1}{2}\mathbb{P}_{2L-2,2L}\mathbb{K}_{2L-1,2L}\right)Q^{B_1}_{2L}\\
&+\lambda^2 Q^{A_1}_1\left(\mathbb{I}-Q_2^{A^{\dg}_1}\right)Q^{B_1}_{2L}+\lambda^2 Q^{A_1}_1\left(\mathbb{I}-Q_{2L-1}^{B^{\dg}_1}\right)Q^{B_1}_{2L},
\end{split}
\end{equation}
where the trace operator $\mathbb{K}$ and permutation operator $\mathbb{P}$ are defined as
\begin{equation}
(\mathbb{K}_{ij})^{I_iI_j}_{J_iJ_j}=\delta^{I_iI_j}\delta_{J_iJ_j},~~~
(\mathbb{P}_{ij})^{I_iI_j}_{J_iJ_j}=\delta^{I_j}_{J_i}\delta_{J_i}^{I_j},
\end{equation}
and the $Q$ operators are defined as
\begin{equation}
Q^{\phi}|\phi\rangle=0,~~~Q^{\phi}|\psi\rangle=|\psi\rangle,~~~\mathrm{for }~\psi \neq \phi.
\end{equation}
\par The shortest operator described the one particle excitation has the form $\ket{p}_{B_1^{\dagger}}$=$f_{B_1^{\dagger}}(1)\ket{1}_{B_1^{\dagger}}+f_{B_1^{\dagger}}(2)\ket{2}_{B_1^{\dagger}}$ with $L=2$, where we have used the same notation in \cite{Chen:2018sbp}. The anomalous dimension of this operator is related to bulk energy $\epsilon(p)$ of the magnon as
\be
\Delta=\epsilon(p)-\frac{1}{2}=\frac{1}{2}\sqrt{1+16h^2(\l)\sin^2(\frac{p}{2})}-\frac{1}{2}=4\lambda^2\sin^2(\frac{p}{2})+\mathcal{O}(\lambda^3).
\ee
The eigenvalue equation is
\be
H\ket{p}_{B_1^{\dagger}}=\Delta\ket{p}_{B_1^{\dagger}}=\matr{cc}
{
\l^2&-\l^2\\
-\l^2&2\l^2\\
}\matr{cc}{f_{B_1^{\dagger}}(1)\\f_{B_1^{\dagger}}(2)}.
\ee
Comparing with the solution of the eigenvalue equation in \cite{Chen:2018sbp}, we can get the first eigenvalue when $f_{B_1^{\dagger}}(1)/f_{B_1^{\dagger}}(2)=\frac{1-\sqrt{5}}{2}$ i.e. when $p=3\pi/5$
\be
\Delta_+=\frac{3+\sqrt{5}}{2}\l^2=4\l^2\sin^2(\frac{3\pi}{10}),
\ee
and the second eigenvalue with $f_{B_1^{\dagger}}(1)/f_{B_1^{\dagger}}(2)=\frac{1+\sqrt{5}}{2}$, which means $p=\frac{\pi}{5}$
\be
\Delta_-=\frac{3-\sqrt{5}}{2}\l^2=4\l^2\sin^2(\frac{\pi}{10}).
\ee
In a similar way, we find the possible momentum values of a single $A_1$ excitation to be $p=\frac{\pi}{5},\frac{3\pi}{5}$ with $L=3$. For
$L=4$, we obtain $p=\frac{\pi}{7},\frac{3\pi}{7},\frac{5\pi}{7}$.\\
\par We now turn to the boundary Bethe-Yang (BBY) equation of a single particle excitation. As mentioned previously, the boundary scattering amplitudes are fixed by
boundary symmetry and boundary crossing as
\be
\mathbb{R}^{+}(-p)=\mathbb{R}^{-}(p)=\mathbb{R}(p)=R_0(p)\diag(e^{-i\frac{p}{2}},-e^{i\frac{p}{2}},1,1),
\ee
where
\be\label{R0}
R_0^2(p)=-e^{-\frac{ip}{2}}\frac{1}{\sqrt{f_b(p)}}.
\ee
Then for a single particle excitation, the BBY equation reads
\be\label{BBY}
e^{-2ipL}\mathbb{R}^{+}(-p)\mathbb{R}^{-}(p)=e^{-2ipL}R_0^2(p)\diag(e^{-ip},e^{ip},1,1)=1.
\ee
At leading order, this reduce to
\be
e^{-2ipL}\diag(e^{-ip},e^{ip},1,1)=-1.
\ee
For a single $B_1^{\dagger}$ excitation, we obtain the quantized momentum
\be
p_n=\frac{n\pi}{2L+1},\qquad n=1,3,\cdots,2L-1.
\ee
Similarly, for a single $A_1$ excitation, the quantized momentum is
\be
p_n=\frac{n\pi}{2L-1},\qquad n=1,3,\cdots,2L-3.
\ee
These give the same results with the Hamiltonian based computation.
\section{Asymptotic Bethe ansatz equaitons}\label{sec4}
\subsection{Double row transfer matrices}
As mentioned before, there are two type of excitations in ABJM open spin chain from giant graviton which we call A-particles and B-particles. The two-body S-matrices between elementary excitations can be written as
\begin{equation}
\begin{split}
\mathbb{S}^{AA}(p_1,p_2)=\mathbb{S}^{BB}(p_1,p_2)=S_0(p_1,p_2)S(p_1,p_2),\\
\mathbb{S}^{AB}(p_1,p_2)=\mathbb{S}^{BA}(p_1,p_2)=\tilde{S}_0(p_1,p_2)S(p_1,p_2),
\end{split}
\end{equation}
where $S(p_1,p_2)$ is the $\mathfrak{su}(2|2)$ invariant S-matrix normalized in $\mathfrak{su}(2)$ compatible way \cite{Arutyunov:2008zt} and we choose the scalar factor in the $\mathfrak{su}(2)$ grading as
\be
S_0(p_1,p_2)=\frac{x_1^+-x_2^-}{x_1^--x_2^+}\frac{1-\frac{1}{x_1^+x_2^-}}{1-\frac{1}{x_1^-x_2^+}}\sqrt{\frac{x_1^-}{x_1^+}}\sqrt{\frac{x_2^+}{x_2^-}}\sigma(p_1,p_2), \qquad\tilde{S}_0(p_1,p_2)=\sqrt{\frac{x_1^-}{x_1^+}}\sqrt{\frac{x_2^+}{x_2^-}}\sigma(p_1,p_2).
\ee
\par In order to obtain the right boundary Bethe-Yang equations from the double row transfer matrices, we should define the fundamental double row transfer matrices as following \cite{Ahn:2000jd, Bajnok:2010ui, Bajnok:2012xc}
\footnote{Fundamental means we trace over the four-dimensional representation of centrally extended $\mathfrak{su}(2|2)$, and we arrange the supersymmetric grading as $BBFF$.}
\be\small{
\mathbb{D}(p,\{p_i^A,p_i^B\})=\Tr_a\lt(\prod_{\scriptscriptstyle{i=K^I}}^{\scriptscriptstyle{K^I_A+1}}\mathbb{S}_{ai}^{AB}(p,p_{\tilde{i}}^B)\prod_{\scriptscriptstyle{i=K^I_A}}^1\mathbb{S}_{ai}^{AA}(p,p_i^A)\mathbb{R}_a^-(p)
\prod_{\scriptscriptstyle{i=1}}^{\scriptscriptstyle{K^I_A}}\mathbb{S}_{ia}^{AA}(p_i^A,-p)\prod_{\scriptscriptstyle{i=K^I_A+1}}^{\scriptscriptstyle{K^I}}\mathbb{S}_{ia}^{AB}(p_{\tilde{i}}^B,-p)\breve{\mathbb{R}}_a^+(-p)\rt)}\nn
\ee
\be\small{
\tilde{\mathbb{D}}(p,\{p_i^A,p_i^B\})=\Tr_{a}\lt(\prod_{\scriptscriptstyle{i=K^I}}^{\scriptscriptstyle{K^I_A+1}}\mathbb{S}_{ai}^{BB}(p,p_{\tilde{i}}^B)\prod_{\scriptscriptstyle{i=K^I_A}}^1\mathbb{S}_{ai}^{BA}(p,p_i^A)\mathbb{R}_a^-(p)
\prod_{\scriptscriptstyle{i=1}}^{\scriptscriptstyle{K^I_A}}\mathbb{S}_{ia}^{BA}(p_i^A,-p)\prod_{\scriptscriptstyle{i=K^I_A+1}}^{\scriptscriptstyle{K^I}}\mathbb{S}_{ia}^{BB}(p_{\tilde{i}}^B,-p)\breve{\mathbb{R}}_a^+(-p)\rt)}\nn
\ee
where
\be
\mathbb{R}^-(p)=R_0(p)R^-(p),\qquad \breve{\mathbb{R}}^+(-p)=\breve{R}^+_0(-p)\breve{R}^+(-p),\qquad K^I=K^I_A+K^I_B,\qquad \tilde{i}=i-K^{I}_A.
\ee
$R_0^-$ is given by eq.~\ref{R0} and
\be
R^-(p)=\diag(e^{-\frac{ip}{2}},-e^{\frac{ip}{2}},1,1).
\ee
$\breve{\mathbb{R}}^+(p)$ is defined through
\be\label{Rbreve}
\mathbb{R}^-_{a}(p)=\Tr_{a'}(\mathbb{P}_{aa'}\mathbb{S}_{aa'}^{AA}(p,-p)\breve{\mathbb{R}}^+_{a'}(-p))
\ee
such that the boundary Bethe-Yang equations can be obtained from the double row transfer matrices as
\be\label{BBY1}
e^{-2ip_j^AL}\mathbb{D}(p_j^A,\{p_i^A,p_i^B\})=-1,\qquad e^{-2ip_j^BL}\tilde{\mathbb{D}}(p_j^B,\{p_i^A,p_i^B\})=-1.
\ee
Using the explicit form of the S-matrix, one can solve the equation \ref{Rbreve} as\footnote{Alternatively, one can determinate the scalar factor $\breve{R}_0(-p)$ by comparing the one particle BBY equation \ref{BBY} and eq.~\ref{BBY1} \cite{Bajnok:2012xc}. Although not shown here, the two methods agree with each other.}
\be\label{Rbreve1}
\breve{R}^+_0(-p)=\frac{e^{-ip}R_0^-(p)}{S_0(p,-p)\rho(p)},\qquad \breve{R}^+(-p)=(-1)^FR^-(-p)=\diag(e^{\frac{ip}{2}},-e^{-\frac{ip}{2}},-1,-1),
\ee
where
\be
\rho=\frac{(1+(x^-)^2)(x^++x^-)}{2x^+(1+x^+x^-)}.
\ee
\par The two kinds of double row transfer matrices defined above actually differ only in some scalar factors. We introduce
\be\label{D}\small{
D(p,\{p_i^A,p_i^B\})=\Tr_a\lt(\prod_{\scriptscriptstyle{i=K^I}}^{\scriptscriptstyle{K^I_A+1}}S_{ai}(p,p_{\tilde{i}}^B)\prod_{\scriptscriptstyle{i=K^I_A}}^1S_{ai}(p,p_i^A)R_a^-(p)
\prod_{\scriptscriptstyle{i=1}}^{\scriptscriptstyle{K^I_A}}S_{ia}(p_i^A,-p)\prod_{\scriptscriptstyle{i=K^I_A+1}}^{\scriptscriptstyle{K^I}}S_{ia}(p_{\tilde{i}}^B,-p)\breve{R}_a^+(-p)\rt)}.\nn
\ee
Then the two kinds double row transfer matrices can be related as
\be
\mathbb{D}(p,\{p_i^A,p_i^B\})=d(p)D(p,\{p_i^A,p_i^B\}),\qquad \tilde{\mathbb{D}}(p,\{p_i^A,p_i^B\})=\tilde{d}(p)D(p,\{p_i^A,p_i^B\}),
\ee
where
\be
\begin{split}
&d(p)=R_0^-(p)\breve{R}^+_0(-p)\prod_{\scriptscriptstyle{i=1}}^{\scriptscriptstyle{K^I_A}}S_0(p,p_i^A)S_0(p_i^A,-p)
\prod_{\scriptscriptstyle{i=1}}^{\scriptscriptstyle{K^I_B}}\tilde{S}_0(p,p_i^B)\tilde{S}_0(p_i^B,-p),\\
&\tilde{d}(p)=R_0^-(p)\breve{R}^+_0(-p)\prod_{\scriptscriptstyle{i=1}}^{\scriptscriptstyle{K^I_A}}\tilde{S}_0(p,p_i^A)\tilde{S}_0(p_i^A,-p)
\prod_{\scriptscriptstyle{i=1}}^{\scriptscriptstyle{K^I_B}}S_0(p,p_i^B)S_0(p_i^B,-p).
\end{split}
\ee
\par Due to the relation \ref{Rbreve1}, we can change the trace in the definition \ref{D} to supertrace
\be\small{
D(p,\{p_i^A,p_i^B\})=\mathrm{sTr}_a\lt(\prod_{\scriptscriptstyle{i=K^I}}^{\scriptscriptstyle{K^I_A+1}}S_{ai}(p,p_{\tilde{i}}^B)\prod_{\scriptscriptstyle{i=K^I_A}}^1S_{1i}(p,p_i^A)R_a^-(p)
\prod_{\scriptscriptstyle{i=1}}^{\scriptscriptstyle{K^I_A}}S_{ia}(p_i^A,-p)\prod_{\scriptscriptstyle{i=K^I_A+1}}^{\scriptscriptstyle{K^{I}}}S_{ia}(p_{\tilde{i}}^B,-p)R_a^-(-p)\rt)}.\nn
\ee
The eigenvalue of the fundamental double row transfer matrix of open spin chain from giant graviton has been conjectured in \cite{Bajnok:2012xc} in the SYM context. However, we just need slightly change some definitions of related functions in that paper for our ABJM case. Taking into account the minor differences, we can write down the double row transfer matrix eigenvalue of ABJM open spin chain from giant graviton explicitly as
\begin{equation}
\begin{split}
\Lambda(x_a)=&\left(\frac{x_a^+}{x_a^-}\right)^{K^{\uppercase\expandafter{\romannumeral2}}}\rho(x_a)\bigg[
\prod_{k=1}^{K^{\uppercase\expandafter{\romannumeral2}}}\frac{x_a^--y_k}{x_a^+-y_k}\frac{y_k+x_a^-}{y_k+x_a^+}\\
&-\prod_{\a=A,B}\prod_{i=1}^{K^{\uppercase\expandafter{\romannumeral1}}_{\a}}\frac{x_{a}^+-x_i^{\a+}}{x_a^+-x_i^{\a-}}\frac{x_i^{\a-}+x_a^+}{x_i^{\a+}+x_a^+}
\prod_{k=1}^{K^{\uppercase\expandafter{\romannumeral2}}}\frac{x_{a}^--y_k}{x_{a}^+-y_k}\frac{y_k+x_a^-}{y_k+x_a^+}
\prod_{l=1}^{K^{\uppercase\expandafter{\romannumeral3}}}\frac{u_a-w_l+i}{u_a-w_l}
\frac{w_l+u_a+i}{w_l+u_a}\\
&-\frac{u_a^+}{u_a^-}\prod_{\a=A,B}\prod_{i=1}^{K^{\uppercase\expandafter{\romannumeral1}}_{\a}}\frac{x_a^+-x_i^{\a+}}{x_a^+-x_i^{\a-}}\frac{x_i^{\a-}+x_a^+}{x_i^{\a+}+x_a^+}
\prod_{k=1}^{K^{\uppercase\expandafter{\romannumeral2}}}\frac{\frac{1}{x_a^+}-y_k}{\frac{1}{x_a^-}-y_k}\frac{y_k+\frac{1}{x_a^+}}{y_k+\frac{1}{x_a^-}}
\prod_{l=1}^{K^{\uppercase\expandafter{\romannumeral3}}}\frac{u_a-w_l-i}{u_a-w_l}\frac{w_l+u_a-i}{w_l+u_a}\\
&+\frac{u_a^+}{u_a^-}\prod_{\a=A,B}\prod_{i=1}^{K^{\uppercase\expandafter{\romannumeral1}}_{\a}}\frac{x_a^+-x_i^{\a+}}{x_a^+-x_i^{\a-}}\frac{x_i^{\a-}+x_a^+}{x_i^{\a+}+x_a^+}
\frac{\frac{1}{x_a^-}-x_i^{\a+}}{\frac{1}{x_a^-}-x_i^{\a-}}\frac{x_i^{\a-}+\frac{1}{x_a^-}}{x_i^{\a+}+\frac{1}{x_a^-}}
\prod_{k=1}^{K^{\uppercase\expandafter{\romannumeral2}}}\frac{\frac{1}{x_a^+}-y_k}{\frac{1}{x_a^-}-y_k}\frac{y_k+\frac{1}{x_a^+}}{y_k+\frac{1}{x_a^-}}\bigg],
\end{split}
\end{equation}
where
\be
x_i^{\alpha\pm}=x^{\pm}(p_i^{\alpha}),\quad \alpha=A,B.
\ee
When we evaluate $\Lambda(p)$ at $p=p_j^{\a}$, only the first term survives,
\be
\Lambda(p_j^{\alpha})=\left(\frac{x_j^{\a+}}{x_j^{\a-}}\right)^{K^{\uppercase\expandafter{\romannumeral2}}}\rho(x_j^{\a})
\prod_{k=1}^{K^{\uppercase\expandafter{\romannumeral2}}}\frac{x_j^{\alpha-}-y_k}{x_j^{\alpha+}-y_k}\frac{y_k+x_j^{\alpha-}}{y_k+x_j^{\alpha+}}.
\ee

\subsection{Bethe ansatz equations}
Bethe equations for the auxiliary roots are obtained from the analytic condition of the double row transfer matrix eigenvalue. There are three types of superficial poles of $\Lambda(x_a)$ at $x_a^+=y_j$,$u_a=w_l$ and $x_a^-=1/y_j$. The analytic constraints give three sets of equations
\footnote{The parameter $v$ is related to $y$ by $v+\frac{1}{v}=y$.}
\be
\begin{split}
\prod_{\alpha=A,B}\prod_{i=1}^{K^{I}_{\alpha}}\frac{y_j-x_i^{\alpha-}}{y_j-x_i^{\alpha+}}\frac{y_j+x_i^{\alpha+}}{y_j+x_i^{\alpha-}}
\prod_{k=1}^{K^{III}}\frac{hv_j-w_k-\frac{i}{2}}{hv_j-w_k+\frac{i}{2}}\frac{hv_j+w_k-\frac{i}{2}}{hv_j+w_k+\frac{i}{2}}=1,\\
\frac{w_l^-}{w_l^+}\prod_{j=1}^{K^{II}}\frac{w_l-hv_j-\frac{i}{2}}{w_l-hv_j+\frac{i}{2}}\frac{w_l+hv_j-\frac{i}{2}}{w_l+hv_j+\frac{i}{2}}\prod_{k=1}^{K^{III}}\frac{w_l-w_k+i}{w_l-w_k-i}\frac{w_l+w_k+i}{w_l+w_k-i}=-1,\\
\prod_{\alpha=A,B}\prod_{i=1}^{K^{I}_{\alpha}}\frac{y_j-x_i^{\alpha+}}{y_j-x_i^{\alpha-}}\frac{y_j+x_i^{\alpha-}}{y_j+x_i^{\alpha+}}\prod_{k=1}^{K^{III}}\frac{hv_j-w_k+\frac{i}{2}}{hv_j-w_k-\frac{i}{2}}\frac{hv_j+w_k+\frac{i}{2}}{hv_j+w_k-\frac{i}{2}}=1.
\end{split}
\ee
We observed that the first and third set of equations are the same.
The main or physical Bethe equations for the massive roots are given by
\be
e^{-2ip_j^AL}d(p_j^A)\Lambda(p_i^A)=-1,\qquad e^{-2ip_j^BL}\tilde{d}(p_j^B)\Lambda(p_j^B)=-1,
\ee
where
\be
d(p_j^A)=\frac{-e^{-ip_j^A/2}\sqrt{f_b(p_j^A)}}{\rho(p_j^A)}e^{-ip_j^AK^{I}}\Sigma(p_j^A)\prod_{k=1}^{K^I_A}\frac{(x_{j}^{A+}-x_{k}^{A-})(1-\frac{1}{x_{j}^{A+}x_{k}^{A-}})}{(x_{j}^{A-}-x_{k}^{A+})(1-\frac{1}{x_{j}^{A-}x_{k}^{A+}})}
\frac{(x_{j}^{A+}+x_{k}^{A+})(1+\frac{1}{x_{j}^{A+}x_{k}^{A+}})}{(x_{j}^{A-}+x_{k}^{A-})(1+\frac{1}{x_{j}^{A-}x_{k}^{A-}})}.\nn
\ee
Here the $\Sigma(p)$ is defined as
\be
\Sigma(p)=\prod_{\a=A,B}\prod_{i=1}^{K^{I}_{\a}}\sigma(p,p_i^{\a})\sigma(p_i^{\a},-p)
\ee
and a very similar expression can be found for $\tilde{d}(p_j^B)$.
\section{Weak coupling limit}\label{sec5}
In order to compare the above asymptotic Bethe equations with the two-loop Bethe equations derived in our previous paper \cite{Bai:2019soy}, we should rewrite the asymptotic Bethe equations into a manifestly $OSp(2,2|6)$ covariant way. In doing so we relabel the roots as
\be
\begin{split}
&x_i^A\leftrightarrow x_{4,j},\; i=1,2,\cdots,K^I_A=K_4 \qquad x_i^B \leftrightarrow x_{\bar 4,i},\; i=1,2,\cdots,K^I_B=K_{\bar4}\\
&y_j \leftrightarrow \frac{1}{x_{1,j}},\; j=1,2,\cdots, K_1,\qquad y_{K_1+j} \leftrightarrow x_{3,j},\;j=1,2,\cdots,K_3 \quad(K^{II}=K_1+K_3)\\
&w_l\leftrightarrow u_{2,l},\; l=1,2,\cdots,K^{III}=K_2
\end{split}
\ee
In weak coupling limit, we have
\be
x^{\pm}\rightarrow \frac{u\pm\frac{i}{2}}{h},\quad f_b(p)\rightarrow e^{-ip},\quad \Sigma(p)\rightarrow 0.
\ee
The asymptotic Bethe ansatz equations then reduce to
\be
\begin{split}
&1=\frac{Q_2^-}{Q^+_2}\bigg|_{u_{1,k}},\\
&-1=\frac{u^-}{u^+}\frac{Q_1^-Q_3^-Q_2^{++}}{Q_1^+Q_3^+Q_2^{--}}\bigg|_{u_{2,k}},\\
&1=\frac{Q_2^-Q_4^+Q_{\bar4}^+}{Q_2^+Q_4^-Q_{\bar4}^-}\bigg|_{u_{3,k}},\\
&1=\left(\frac{u-\frac{i}{2}}{u+\frac{i}{2}}\right)^{2L'}\frac{Q_4^{++}Q_3^-}{Q_4^{--}Q_3^+}\bigg|_{u_{4,k}},\\
&1=\left(\frac{u-\frac{i}{2}}{u+\frac{i}{2}}\right)^{2L'}\frac{Q_{\bar4}^{++}Q_3^-}{Q_{\bar4}^{--}Q_3^+}\bigg|_{u_{\bar4,k}},
\end{split}
\ee
where $L'=L+\frac{K_1-K_3}{2}+\frac{K_4+K_{\bar4}-1}{2}$ and we have used the common definition of Baxter polynomial
\be
Q_l(u)=\prod_{j=1}^{K_l}(u-u_{l,j})(u+u_{l,j}).
\ee
\subsection{Reducing to the scalar sector}
In order to obtain the scalar sector Bethe equations and to compare with our previous result derived in \cite{Bai:2019soy}, we must do fermionic duality on the Bethe ansatz equations \cite{Beisert:2005di,Bai:2016pxs}. For this purpose, we define
\be
\begin{split}
P_1(x)&=\prod_{j=1}^{K_2}(x-u_{2,j}-\frac{i}{2})(x+u_{2,j}-\frac{i}{2})-\prod_{j=1}^{K_2}(x-u_{2,j}+\frac{i}{2})(x+u_{2,j}+\frac{i}{2})\\
&=-2iK_2x^{2K_2-1}+\cdots.
\end{split}
\ee
The degree of the polynomial $P_1(x)$ is $2K_2-1$ and have $2K_1$ obvious zeros $\{\pm u_{1,k}\}_{k=1,2,\cdots,K_1}$ and 0\footnote{The polynomial $P_1(x)$ defined here and $P_2(x)$ defined below are odd under reflection: $P_i(x)=-P_i(-x),i=1,2$.}. We can write
\be
P_1(x)=\alpha_1 x\prod_{k=1}^{K_1}(x-u_{1,k})(x+u_{1,k})\prod_{k=1}^{\tilde{K_1}}(x-\tilde{u}_{1,k})(x+\tilde{u}_{1,k}),
\ee
where $\tilde{K_1}=K_2-K_1-1,\a_1=-2iK_2$.
Thus we can compute $P_1(u_{2,k}+\frac{i}{2}))/P_1(u_{2,k}-\frac{i}{2})$ in two ways
\be
\begin{split}
&\frac{P_1(u_{2,k}+\frac{i}{2})}{P_1(u_{2,k}-\frac{i}{2})}=-\prod_{j=1}^{K_2}\frac{(u_{2,k}-u_{2,j}+i)(u_{2,k}+u_{2,j}+i)}{(u_{2,k}-u_{2,j}-i)(u_{2,k}+u_{2,j}-i)}\\
&=\frac{u_{2,k}+\frac{i}{2}}{u_{2,k}-\frac{i}{2}}\prod_{j=1}^{K_1}\frac{(u_{2,k}-u_{1,j}+\frac{i}{2})(u_{2,k}+u_{1,j}+\frac{i}{2})}{(u_{2,k}-u_{1,j}-\frac{i}{2})(u_{2,k}-u_{1,j}-\frac{i}{2})}
\prod_{j=1}^{\tilde{K}_1}\frac{(u_{2,k}-\tilde{u}_{1,j}+\frac{i}{2})(u_{2,k}+\tilde{u}_{1,j}+\frac{i}{2})}{(u_{2,k}-\tilde{u}_{1,j}-\frac{i}{2})(u_{2,k}-\tilde{u}_{1,j}-\frac{i}{2})}.
\end{split}
\ee
In terms of Baxter polynomial, we have
\be
\frac{Q_2^{++}Q_1^-}{Q_2^{--}Q_1^+}\bigg|_{u_{2,k}}=-\frac{u^+}{u^-}\frac{Q_{\tilde{1}}^+}{Q_{\tilde{1}^-}}\bigg|_{u_{2,k}},
\ee
where $Q_{\tilde{1}}$ is dual Baxter polynomial of the first type
\be
Q_{\tilde{l}}(u)=\prod_{j=1}^{\tilde{K}_l}(u-\tilde{u}_{l,j})(u+\tilde{u}_{l,j}).
\ee
Therefore, after apply fermionic duality on the first Dykin node, we obtain
\be
\begin{split}
&1=\frac{Q_2^+}{Q^-_2}\bigg|_{\tilde{u}_{1,k}},\\
&1=\frac{Q_{\tilde{1}}^+Q_3^-}{Q_{\tilde{1}}^-Q_3^+}\bigg|_{u_{2,k}},\\
&1=\frac{Q_2^-Q_4^+Q_{\bar4}^+}{Q_2^+Q_4^-Q_{\bar4}^-}\bigg|_{u_{3,k}},\\
&1=\left(\frac{u-\frac{i}{2}}{u+\frac{i}{2}}\right)^{2\tilde{L'}}\frac{Q_4^{++}Q_3^-}{Q_4^{--}Q_3^+}\bigg|_{u_{4,k}},\\
&1=\left(\frac{u-\frac{i}{2}}{u+\frac{i}{2}}\right)^{2\tilde{L'}}\frac{Q_{\bar4}^{++}Q_3^-}{Q_{\bar4}^{--}Q_3^+}\bigg|_{u_{\bar4,k}},
\end{split}
\ee
where $2\tilde{L'}=2L'-2K_1+K_2-1$.
\par Then we apply fermionic dual on the second Dykin node in the new basis by defining\footnote{Fermionic duality on Bethe equations is related to the odd Weyl reflection on simple root systems of super Lie algebra. See Appendix C of \cite{Bai:2016pxs} for more details.}
\be
\begin{split}
P_2(x)&=\prod_{j=1}^{\tilde{K}_1}(x-\tilde{u}_{1,j}+\frac{i}{2})(x+\tilde{u}_{1,j}+\frac{i}{2})
\prod_{l=1}^{K_3}(x-u_{3,l}-\frac{i}{2})(x+u_{3,l}-\frac{i}{2})\\
&-\prod_{j=1}^{\tilde{K}_1}(x-\tilde{u}_{1,j}-\frac{i}{2})(x+\tilde{u}_{1,j}-\frac{i}{2})
\prod_{l=1}^{K_3}(x-u_{3,l}+\frac{i}{2})(x+u_{3,l}+\frac{i}{2})\\
&=-2i(K_3-\tilde{K}_1)x^{2\tilde{K}_1+2K_3-1}+\cdots.
\end{split}
\ee
The degree of the polynomial $P_2(x)$ is $2\tilde{K}_1+2K_3-1$ and have $2K_2$ obvious zeros $\{\pm u_{2,k}\}_{k=1,2,\cdots,K_2}$ and 0. We can write
\be
P_2(x)=\alpha_2 x\prod_{k=1}^{K_2}(x-u_{2,k})(x+u_{2,k})\prod_{k=1}^{\tilde{K_2}}(x-\tilde{u}_{2,k})(x+\tilde{u}_{2,k}),
\ee
where $\tilde{K_2}=\tilde{K}_1+K_3-K_2-1=K_3-K_1-2,\a_2=-2i(K_3-\tilde{K}_1)$.
Similarly, by using the two different expression of $P_2(x)$, we can get the following relations
\be
\begin{split}
&\frac{P_2(u_{3,k}+\frac{i}{2})}{P_2(u_{3,k}-\frac{i}{2})}=-\prod_{j=1}^{K_3}\frac{(u_{3,k}-u_{3,l}+i)(u_{3,k}+u_{3,l}+i)}{(u_{3,k}-u_{3,l}-i)(u_{3,k}+u_{3,l}-i)}\\
&=\frac{u_{3,k}+\frac{i}{2}}{u_{3,k}-\frac{i}{2}}\prod_{j=1}^{K_2}\frac{(u_{3,k}-u_{2,j}+\frac{i}{2})(u_{3,k}+u_{2,j}+\frac{i}{2})}{(u_{3,k}-u_{2,j}-\frac{i}{2})(u_{3,k}-u_{2,j}-\frac{i}{2})}
\prod_{j=1}^{\tilde{K}_2}\frac{(u_{3,k}-\tilde{u}_{2,j}+\frac{i}{2})(u_{3,k}+\tilde{u}_{2,j}+\frac{i}{2})}{(u_{3,k}-\tilde{u}_{2,j}-\frac{i}{2})(u_{3,k}-\tilde{u}_{2,j}-\frac{i}{2})}
\end{split}
\ee
and
\be
\begin{split}
&\frac{P_2(\tilde{u}_{1,k}+\frac{i}{2})}{P_2(\tilde{u}_{1,k}-\frac{i}{2})}=-\prod_{j=1}^{\tilde{K}_1}\frac{(\tilde{u}_{1,k}-\tilde{u}_{1,j}+i)(\tilde{u}_{1,k}+\tilde{u}_{1,j}+i)}{(\tilde{u}_{1,k}-\tilde{u}_{1,j}-i)(\tilde{u}_{1,k}+\tilde{u}_{1,j}-i)}\\
&=\frac{\tilde{u}_{1,k}+\frac{i}{2}}{\tilde{u}_{1,k}-\frac{i}{2}}\prod_{j=1}^{K_2}\frac{(\tilde{u}_{1,k}-u_{2,j}+\frac{i}{2})(\tilde{u}_{1,k}+u_{2,j}+\frac{i}{2})}{(\tilde{u}_{1,k}-u_{2,j}-\frac{i}{2})(\tilde{u}_{1,k}-u_{2,j}-\frac{i}{2})}
\prod_{j=1}^{\tilde{K}_2}\frac{(\tilde{u}_{1,k}-\tilde{u}_{2,j}+\frac{i}{2})(\tilde{u}_{1,k}+\tilde{u}_{2,j}+\frac{i}{2})}{(\tilde{u}_{1,k}-\tilde{u}_{2,j}-\frac{i}{2})(\tilde{u}_{1,k}-\tilde{u}_{2,j}-\frac{i}{2})}.
\end{split}
\ee
Using the above relations, the following Bethe equations after dualization can be easily derived
\be
\begin{split}
&-1=\frac{u^{-}}{u^{+}}\frac{Q_{\tilde{1}}^{++}Q_{\tilde{2}}^-}{Q_{\tilde{1}}^{--}Q^+_{\tilde{2}}}\bigg|_{\tilde{u}_{1,k}},\\
&1=\frac{Q_{\tilde{1}}^-Q_3^+}{Q_{\tilde{1}}^+Q_3^-}\bigg|_{\tilde{u}_{2,k}},\\
&-1=\frac{u^+}{u^-}\frac{Q_{\tilde{2}}^+Q_3^{--}Q_4^+Q_{\bar4}^+}{Q_{\tilde{2}}^-Q_3^{++}Q_4^-Q_{\bar4}^-}\bigg|_{u_{3,k}},\\
&1=\left(\frac{u-\frac{i}{2}}{u+\frac{i}{2}}\right)^{2\tilde{L'}}\frac{Q_4^{++}Q_3^-}{Q_4^{--}Q_3^+}\bigg|_{u_{4,k}},\\
&1=\left(\frac{u-\frac{i}{2}}{u+\frac{i}{2}}\right)^{2\tilde{L'}}\frac{Q_{\bar4}^{++}Q_3^-}{Q_{\bar4}^{--}Q_3^+}\bigg|_{u_{\bar4,k}}.
\end{split}
\ee
\par We now should remove the first and second type of Bethe roots $\tilde{u}_{1,k},\tilde{u}_{2,k}$ to obtain the scalar sector Bethe equation. Applying ``gauge" transformation: $Q_3(u)\rightarrow u^2Q_3(u)$ and identifying $\tilde{L'}=L$ \cite{Nepomechie:2009zi,Nepomechie:2011nz},
we found it has the same form with the equations given in \cite{Bai:2019soy}. See appendix \ref{appenA} for details.
\section{Conclusion}\label{sec6}
In this paper, we have derived the all loop Bethe ansatz equations for our ABJM open spin chain constructed from giant graviton mainly based on symmetry analysis. We check our result in the weak coupling region by comparing with the two-loop $SU(4)$ sector Bethe ansatz equations given in our previous work. By using fermionic duality and ``gauge" transformation, we found our proposal in this paper is consistent with our previous results. It's interesting to go beyond the asymptotic region to include all finite-size effect in the boundary thermodynamic Bethe ansatz (BTBA) framework of the ABJM open spin chain from giant graviton. The similar treatment of integrable open system from giant graviton in SYM is fruitful and we hope this also happens in ABJM theory.
\section*{Acknowledgments}
I would like to thank Jun-Bao Wu, Hao Ouyang and Rafael I. Nepomechie for very helpful discussions. I am also grateful to Changrim Ahn for very valuable correspondence at an early stage of this work.
\begin{appendix}
\section{``Gauge" transformation of Bethe equations}\label{appenA}
In this appendix, we show that the two-loop scalar sector Bethe ansatz equations have ``gauge" freedom, i.e. the form of Bethe equations is not unique. The different forms are related by ``gauge" transformation which we now discuss. As computed in \cite{Bai:2019soy}, at two-loop orders, the vacuum eigenvalues of double row transfer matrices are given by \footnote{The two-loop construction of double row transfer matrices is described in detail in paper \cite{Bai:2019soy}. In that paper, only one possible eigenvalues of double row transfer matrices were given. In this appendix, we give an alternative one and find relations of these two solutions.}
\begin{equation}
\Lambda_0(u)=\bar\Lambda_0(u)=\frac{2}{d(u)}\Big[a(u)(u+1)^{2L} (u+2)^{2L}+b(u)u^{2L}(u+1)^{2L}-c(u)u^{2L}(u+2)^{2L}\Big],
\end{equation}
where
\be
\begin{split}
a(u)=(2u+3)(u+1)^2,\quad b(u)=(2u+1)(u+1)^2,\\
c(u)=4(u+1)^3,\quad d(u)=(u+1)(2u+1)(2u+3).
\end{split}
\ee
The eigenvalues of a generic state should have the ``dressed" form
\be\label{L}
\begin{split}
\Lambda(u|\{u_i\})= &\frac{2}{d(u)}\Big\{a(u)(u+1)^{2L} (u+2)^{2L}\frac{Q_4(iu-\frac{i}{2})}{Q_4(iu+\frac{i}{2})}
+b(u)u^{2L}(u+1)^{2L}\frac{Q_{\bar4}(iu+\frac{5i}{2})}{Q_{\bar4}(iu+\frac{3i}{2})}\\
&-u^{2L}(u+2)^{2L}\Big[c_1(u)\frac{Q_4(iu+\frac{3i}{2})Q_3(iu)}{Q_4(u+\frac{i}{2})Q_3(iu+i)}
+c_2(u)\frac{Q_3(iu+2i)Q_{\bar4}(iu+\frac{i}{2})}{Q_3(iu+i)Q_{\bar4}(iu+\frac{3i}{2})}\Big]\Big\},
\end{split}
\ee
\be\label{Lb}
\begin{split}
\bar\Lambda(u|\{u_i\})= &\frac{2}{d(u)}\Big\{a(u)(u+1)^{2L} (u+2)^{2L}\frac{Q_{\bar4}(iu-\frac{i}{2})}{Q_{\bar4}(iu+\frac{i}{2})}
+b(u)u^{2L}(u+1)^{2L}\frac{Q_{4}(iu+\frac{5i}{2})}{Q_{4}(iu+\frac{3i}{2})}\\
&-u^{2L}(u+2)^{2L}\Big[c_1(u)\frac{Q_{\bar4}(iu+\frac{3i}{2})Q_3(iu)}{Q_{\bar4}(u+\frac{i}{2})Q_3(iu+i)}
+c_2(u)\frac{Q_3(iu+2i)Q_{4}(iu+\frac{i}{2})}{Q_3(iu+i)Q_{4}(iu+\frac{3i}{2})}\Big]\Big\}.
\end{split}
\ee
The functions $c_1(u),c_2(u)$ must satisfy
\be\label{ceq1}
c_1(u)+c_2(u)=c(u).
\ee
The crossing property of eigenvalues
\be
\Lambda(-u-2|\{u_i\})=\bar\Lambda(u|\{u_i\})
\ee
implies
\be\label{ceq2}
c_1(-u-2)=-c_2(u),\quad c_2(-u-2)=-c_1(u).
\ee
The constraints eq.~\ref{ceq1} and eq.~\ref{ceq2} cannot determine $c_1(u),c_2(u)$ uniquely. In fact, there are two solutions
\be
c_1(u)=(2u+3)(u+1)^2,\quad c_2(u)=(u+1)^2(2u+1),
\ee
and
\be
\tilde{c}_1(u)=u^2(2u+3),\quad \tilde{c}_2(u)=(u+2)^2(2u+1).
\ee
In terms of the eigenvalues of double row transfer matrices eq.~\ref{L} and eq.~\ref{Lb}, these two solutions can be related by the gauge transformation on the Baxter polynomial $Q_3(u)$
\be
Q_3(u)\rightarrow u^2Q_3(u).
\ee
%
%\be
%\begin{split}
% \bar\Lambda(u|\{u_i\})=  & \frac{2u^{2L}(u+1)^{2L+1}}{2u+3}\frac{Q_1(u+\frac52)}{Q_1(u+\frac32)}
% -\frac{2u^{2L}(u+2)^{2L+2}}{(2u+3)(u+1)}
%\frac{Q_1(u+\frac12)Q_2(u+2)}{Q_1(u+\frac32)Q_2(u+1)}
%\\
%&-\frac{2u^{2L+2}(u+2)^{2L}}{(2u+1)(u+1)}\frac{Q_3(u+\frac32)Q_2(u)}{Q_3(u+\frac12)Q_2(u+1)}
%+\frac{2(u+1)^{2L+1}(u+2)^{2L}}{2u+1}\frac{Q_3(u-\frac12)}{Q_3(u+\frac12)}
%\end{split}
%\ee
\end{appendix}

\end{document}